\documentclass[aps,pra,twocolumn,twoside,superscriptaddress,notitlepage]{revtex4-2}
\usepackage{tikz}
\usetikzlibrary{calc}
\usetikzlibrary{shapes.geometric, arrows}
\usepackage{eufrak}
\usepackage{enumitem}
\usepackage{mathtools}
\usepackage{graphicx}
\usetikzlibrary{positioning}
\usepackage{bm}
\newcommand{\bematrix}{\left(\begin{matrix}}
\newcommand{\ematrix}{\end{matrix}\right)}

\usepackage{ulem} 
\usepackage[caption=false]{subfig}
\usepackage{dcolumn}
\usepackage{amsmath,amssymb}
\usepackage{bm}
\usepackage{bbm}
\usepackage{overpic}
\usepackage{latexsym}
\usepackage{color}
\usepackage[english]{babel}
\usepackage{latexsym}
\usepackage{psfrag,graphicx}
\usepackage{epsf}
\usepackage{amsmath}
\usepackage{amssymb}
\usepackage{amsfonts}
\usepackage{natbib}
\usepackage{multirow} 
\usepackage{appendix}
\usepackage{verbatim}
\usepackage{enumitem}
\usepackage{dsfont}

\usepackage{float}
\usepackage{tikz}
\DeclareMathOperator{\aux}{aux}
\DeclareMathOperator{\QFT}{QFT}
\DeclareMathOperator{\T}{T}
\definecolor{mygrey}{gray}{0.35}
\definecolor{myblue}{rgb}{0.2,0.2,0.8}
\definecolor{myzard}{cmyk}{0,0,0.05,0}
\definecolor{mywhite}{rgb}{1,1,1}
\definecolor{myred}{rgb}{0.9,0.1,0.}
\usepackage[colorlinks=true,citecolor=myblue,linkcolor=myblue,urlcolor=myblue]{hyperref}

\usepackage[makeroom]{cancel}

\DeclareMathOperator{\Tr}{Tr}

\newenvironment{proof-of}[1]{\medskip\noindent\textbf{Proof of {#1}.}}{\hfill$\blacksquare$\medskip}
\newcommand{\ket}[1]{\left\vert#1\right\rangle}
\newcommand{\bra}[1]{\left\langle#1\right\vert}
\newcommand{\braket}[2]{\ensuremath{\langle #1 | #2 \rangle}}

\usepackage{colortbl}
\usepackage{xcolor}

\usepackage{tcolorbox}

\definecolor{lightgray}{gray}{0.9}

\begin{document}

\title{Pre-detection squeezing as a resource for high-dimensional Bell-state measurements}

\author{Luca Bianchi}
\affiliation{Department of Physics and Astronomy, University of Florence, 50019, Firenze, Italy}

\author{Carlo Marconi}
\affiliation{Istituto Nazionale di Ottica del Consiglio Nazionale delle Ricerche (CNR-INO), 50125 Firenze, Italy}

\author{Jan Sperling}
\affiliation{Theoretical Quantum Science, Institute for Photonic Quantum Systems (PhoQS), Paderborn University, Warburger Stra\ss{}e 100, 33098 Paderborn, Germany}

\author{Davide Bacco}
\email{davide.bacco@unifi.it}
\affiliation{Department of Physics and Astronomy, University of Florence, 50019, Firenze, Italy}

\begin{abstract}
    Bell measurements, entailing the projection onto one of the Bell states, play a key role in quantum information and communication, where the outcome of a variety of protocols crucially depends on the success probability of such measurements.
    Although in the case of qubit systems, Bell measurements can be implemented using only linear optical components, the same result is no longer true for qudits, where at least the use of ancillary photons is required.
    In order to circumvent this limitation, one possibility is to introduce nonlinear effects.
    In this work, we adopt the latter approach and propose a scalable Bell measurement scheme for high-dimensional states, exploiting multiple squeezer devices applied to a linear optical circuit for discriminating the different Bell states. 
    Our approach does not require ancillary photons, is not limited by the dimension of the quantum states, and is experimentally scalable, thus paving the way toward the realization of an effective high-dimensional Bell measurement.
\end{abstract}

\date{\today}

\maketitle

\section{Introduction}

    With the advent of quantum information, computational and communication tasks are expected to be executed using devices governed by the principles of quantum mechanics \cite{preskill2023quantum}. Therefore, it is of high importance to establish a practical way to link quantum nodes through a quantum network \cite{chen2021review, wehner2018quantum, ribezzo2023deploying, kimble2008quantum, pirandola2019end, pirandola2020advances}.
    In this sense, quantum optical platforms, where the information is stored in photonic degrees of freedom, appear to be promising candidates for long-range transmissions \cite{xu2020secure}.
    In fact, photons are characterized by low decoherence and absence of self-interactions and can be easily deployed via readily available optical fiber networks \cite{slussarenko2019photonic}.
    However, one of the main drawbacks of this approach is represented by detrimental phenomena, such as photon losses and absorption, that inevitably affect the propagation of quantum systems through optical fibers \cite{schlosshauer2004decoherence, scarani2009security}.
    
    To mitigate such effects, quantum repeaters, based on different technologies, have been proposed and proof-of-concept experiments have already been implemented \cite{azuma2023quantum, pirandola2015general}. However, most of the current schemes are based on qubit encodings, which are limited in terms of photon information efficiency and noise resilience\cite{cozzolino2019high}. 
    
    In contrast, qudits, that is, high-dimensional systems, \cite{cozzolino2019high,wang2020qudits} allow higher photon information capacity, higher resistance to noise, and improved cryptographic security \cite{Dixon, Xiao, barreiro2008beating}.
    In recent years, qudits have been successfully encoded and transmitted on a variety of different platforms \cite{cozzolino2019orbital,islam2017provably,kues2017chip,krenn2017entanglement,llewellyn2020chip, zahidy2024practical}.
    In particular, path encoding has been proven to be effective for the generation, manipulation, and detection of photonic qudits, as well as for noise resilience \cite{da2021path}.

    One of the simplest models for a quantum repeater is based on the entanglement swapping protocol \cite{entanglementswapping}, whose fundamental ingredient consists of a Bell-state measurement (BSM) \cite{braunstein1995measurement, pirandola2015advances}.
    Such a task encompasses the projection onto one of the Bell states and is fundamental to entangle initially uncorrelated particles. In the case of two qubits, a BSM can be implemented by a balanced beam splitter and photodetectors placed at its output ports \cite{bouwmeester1997experimental}.
    Through Hong-Ou-Mandel interference \cite{HOM}, distinguishable input photon pairs can be uniquely associated with given Bell states, while indistinguishable photons give an ambiguous discrimination. 
    This means that this scheme allows to distinguish only two among the four Bell states, thus yielding a success probability of 50\% for Bell state discrimination. Interestingly, this corresponds also the to the maximum success probability attainable for qubit systems in linear optics \cite{lutkenhaus1999bell}.
    Unfortunately, in the case of high-dimensional systems, it was proven that, under the same restriction to use only linear optical components, it is impossible to perform a BSM\cite{carollo2001role, calsamiglia2002generalized}.
    Such no-go result can be circumvented introducing auxiliary input states \cite{duvsek2001discrimination}, although the implementation of such states is experimentally challenging and highly impractical for the purpose of a scalable repeater architecture.

    Other proposals have been investigated both for qubit and qudit systems. For qubits, there exist several proposals based, e.g., on hyper-entanglement \cite{walborn2003hyperentanglement}, ancillary photons \cite{olivo2018ancilla, ewertancilla}, and non-linear optical \cite{kimNLO}, while for high-dimensional systems, a linear optical scheme using auxiliary single photons or other quantum states was proposed for arbitrary dimensions \cite{bacco2021proposal} and experimentally demonstrated in the case of qutrits \cite{luo2019quantum}.
    Nonlinear optical proposals were formulated for orbital-angular-momentum-encoded qudits \cite{forbes}, and a more general one has been recently proposed \cite{bianchi2024nonlinear}. Both proposals show a promising alternative measurement for teleportation and entanglement swapping, but require efficient non linear methods.

    In our work, we exploit the technique based on squeezing introduced in \cite{zaidi2013beating} to propose and demonstrate a scalable solution for the case of high-dimensional systems and show that it allows one to beat the current highest success probability reported in \cite{luo2019quantum} for the case of qutrits.
    Specifically, we first provide an analytic description of our setup using the generating function formalism, in the context of positive operator-valued measures (POVMs) describing photodetectors \cite{sperling2014quantum, sperling2018quasistates}.
    Such a description turns out to be useful whenever it comes to computing quantum-statistical moments \cite{engelkemeier2020quantum} and other related quantities. Then, we compute the success probability of BSM for photonic qudits up to dimension five, thus benchmarking the scalability of our proposal.
    In particular, we find that the pre-detection squeezing method beats the current limitation on the success probability given by linear optics and auxiliary single photons.
    Our proposal does not need the use of additional photons, requires an experimentally feasible amount of squeezing, and can be scaled to the case of arbitrary dimension, thus paving the way towards a practical implementation of a quantum optical repeater.

    The paper is structured as follows.
    In Sec. \ref{2}, we review the current proposals for both qubit (Sec. \ref{2A}) and qudit (Sec. \ref{2B}) state-of-the-art BSMs, noticing the presence of auxiliary photons and the requirement of an higher degree of mixing, as well as the introduction of an additional transformation.
    In Sec. \ref{3}, we describe the methods of our work, particularly introducing the formalism for the analytical description of the POVM of BSM in Sec. \ref{3A}, and we explain how squeezing can be a resource of enhanced interference for the Bell-state discrimination problem.
    In Sec. \ref{3B}, we discuss the details of our numerical simulations.
    Section \ref{4} contains the main results, namely the plots presenting the regions in which the high-dimensional pre-detection squeezing scheme outperforms the auxiliary photon-aided linear optical BSM and the simulations of different dimensions.
    We conclude with Sec. \ref{5} and \ref{6} by commenting on the scalability of our proposal, the possible drawbacks of the scheme, and the prospects of future investigations.

\section{The Bell state measurement}
\label{2}

\subsection{The qubit case}
\label{2A}

    Here we briefly review the scheme to implement a BSM for qubits using linear optical components \cite{braunstein1995measurement}.
    In what follows, we consider the well-known dual-rail encoding \cite{chuang1995simple}, meaning that the logical states for the photonic system are given by $\ket{0_{L}} = \ket{10}$ and $\ket{1_{L}} = \ket{01}$, where zeros and ones on the right-hand side are Fock states for the possible modes (or paths) occupied by the input photons.
    Thus, the Hilbert space of the system, $\mathcal{H} = \mathcal{H}_A \otimes \mathcal{H}_B$, is spanned by the computational basis
    \begin{equation}
    \begin{split}
        \ket{0_L0_L}_{AB} = \ket{1010}_{AB},\quad&
        \ket{0_L1_L}_{AB} = \ket{1001}_{AB},\\
        \ket{1_L0_L}_{AB} = \ket{0110}_{AB},\quad&
        \ket{1_L1_L}_{AB} = \ket{0101}_{AB}.
    \end{split}
    \end{equation}
    As a consequence, the Bell basis can be written as
    \begin{equation}
    \begin{aligned}
        \ket{\Phi^{\pm}}_{AB} &= \frac{1}{2}(\ket{1010}_{AB} \pm \ket{0101}_{AB}),\\
        \ket{\Psi^{\pm}}_{AB} &= \frac{1}{2}(\ket{1001}_{AB} \pm \ket{0110}_{AB}).
    \end{aligned}
    \end{equation}

\begin{figure}
    \centering
    \includegraphics[width=0.5\textwidth]{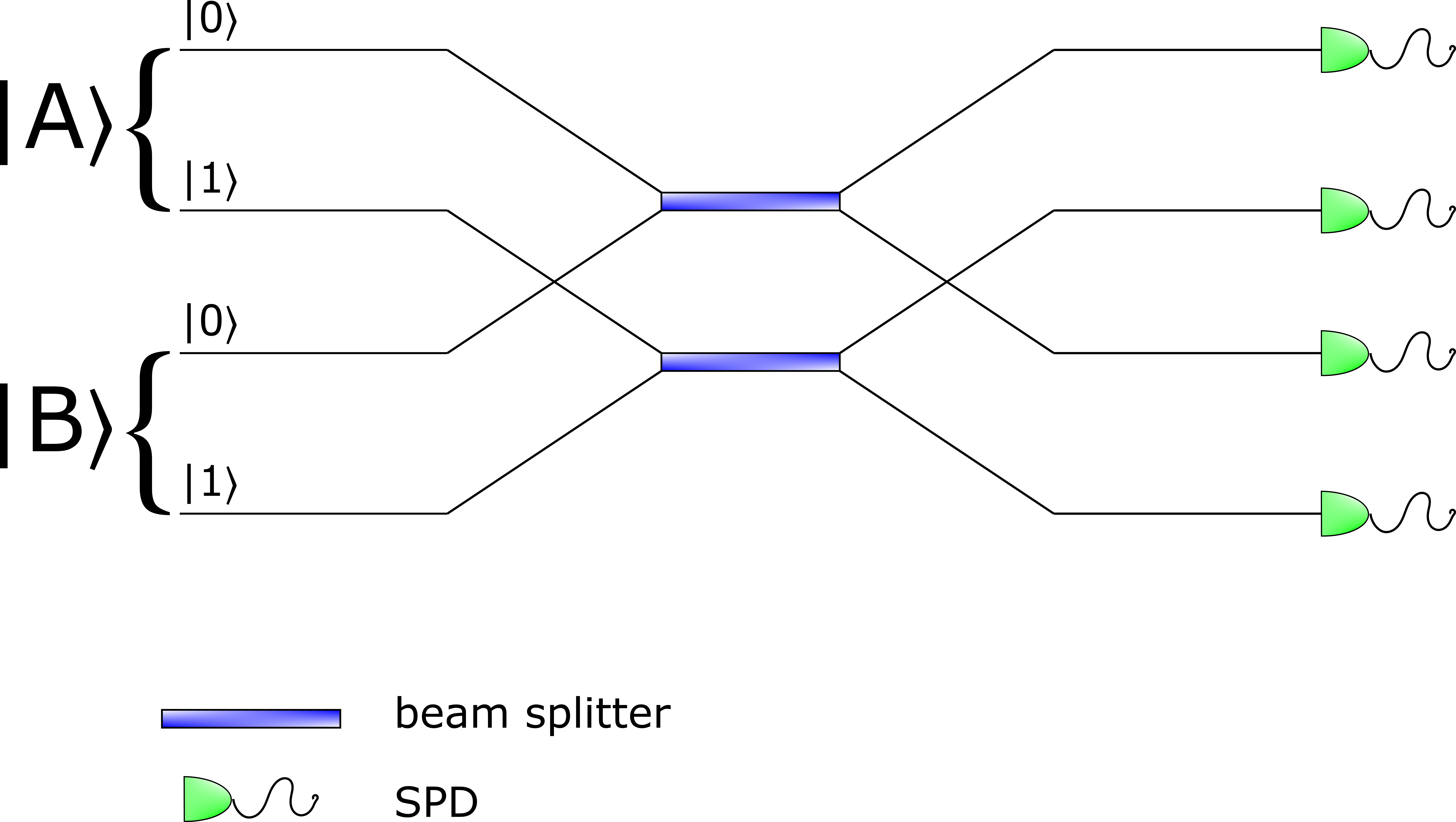}
    \caption{%
        A common BSM for qubits.
        The indistinguishability of photons, guaranteed by mixing the modes inside the beam splitter, yields a BSM with a success probability of $P_s = \frac{1}{2}$.
    }\label{usualscheme}
\end{figure}

    A common scheme for a two-dimensional BSM in dual-rail encoding workes as follows:
    (i) the input Bell states are sent through two $50{:}50$ beam splitters;
    (ii) photodetectors are placed at the output arms of each beam splitter to count the number of photons in each mode;
    see Fig. \ref{usualscheme}.
    As a result of the scheme, each initial Bell state corresponds to a superposition of different terms, associated to certain click patterns.
    Some of these patterns belong uniquely to given input states, while others are ambiguous.
    More specifically, one can see that the states $\ket{\Psi^{\pm}}_{AB}$ can always be unambiguously discriminated while $\ket{\Phi^{\pm}}_{AB}$ are always ambiguous.
    In other words, we say that the states $\ket{\Phi^{\pm}}_{AB}$ are degenerate for the POVM operator associated to the BSM. Still, we can quantify the degree of distinguishability of the Bell states by introducing a figure of merit dubbed success probability, $P_s$, which is calculated by computing the time evolution of the states and then summing the squares of the amplitudes that are uniquely associated to one and only one state.
    It can be seen \cite{lutkenhaus1999bell} that, for a generic linear optical BSM that does not involve auxiliary modes, the success probability is bounded by $P_s = 1/2$.

\subsection{The qudit case}
\label{2B}

    The previous BSM can be extended to high-dimensional systems by raising the number of possible rails that a single photon can take, i.e., by path-encoding the qudits into photons.
    In this scenario, a Bell basis is given by \cite{sych2009complete}
    \begin{equation}
        \ket{\Psi_{lm}} = \frac{1}{\sqrt{d}}\sum_{k=0}^{d-1}\omega_d^{lk}\ket{k_L}\ket{(k\oplus m)_L},
        \label{bell}
    \end{equation}
    where $\oplus$ denotes the sum modulo $d$ and $\omega_d=\exp(2\pi i/d)$.
    Here, the $k$\textsuperscript{th} logical level is identified by a photon occupying the $k$\textsuperscript{th} path, i.e.,
    \begin{equation}
        \ket{k_L} = \ket{0,\dots,1_k,\dots,0}.
    \end{equation}
    However, when the dimension of the logical space is raised, linear optics is not enough to implement an unambiguous BSM.
    In fact, as proven in Ref. \cite{calsamiglia2002generalized}, even probabilistic Bell-state discrimination among any two elements of the qudit Bell basis is impossible using only linear optical transformations.
    In particular, the use of auxiliary photons is required in order to have a POVM set whose rank matches the Schmidt rank of the input states.
    This implies that, for $d$-dimensional Bell states, $d-2$ auxiliary photons are required.
    However, even with the help of extra-modes, a theorem from Carollo and Palma \cite{carollo2001role} states that adding ancillary modes can only boost the probability of performing an incomplete measurement, but it never results in an unambiguous Bell discrimination.

    In Ref. \cite{luo2019quantum}, a scheme for a high-dimensional BSM using ancillary qudits was presented.
    An uncorrelated input pair of qudits $A$ and $B$ is fed into a quantum Fourier transform, described by
    \begin{equation}
        \QFT_d = \frac{1}{\sqrt{d}}
        \begin{pmatrix}
            1 & 1 & 1 & \dots & 1 \\
            1 & \omega_d & \omega_d^2 & \dots & \omega_d^{d-1} \\
            \vdots & \vdots & \vdots & \ddots &  \vdots\\
            1 & \omega_d^{d-1} & \omega_d^{2(d-1)} & \dots & \omega_d^{(d-1)^2}
        \end{pmatrix},
    \end{equation}
    acting on the photon space.
    The remaining $d-2$ input ports of the Fourier network are fed with auxiliary qudits in an equal superposition of modes, i.e.,
    \begin{equation}
        \ket{\aux} = \bigotimes_{x} \frac{1}{\sqrt{d}}\sum_{i=0}^{d-1}\ket{i_L}_{x}.
    \end{equation}
    Furthermore, a transformation $\T_d$---acting before the quantum Fourier transform on the input qudit $B$---is encoded in an extended unitary operation.
    This is required in order to guarantee unambiguous discrimination, and this map takes the form
    \begin{equation}
        \T_d = \frac{1}{d-1}\begin{pmatrix}
            2-d & 1 & \dots & 1 \\
            1 & 2-d & \dots & 1 \\
            \vdots & \vdots & \ddots & \vdots \\
            1 & 1 & \dots & 2-d
        \end{pmatrix}.
    \end{equation}
    At the output ports of the Fourier network, photon-number-resolving detectors are placed--- even if the Schmidt rank projections are given only by specific patterns for single photons detected on different modes.
    In fact, for the $d=3$ case, it can be shown that three click patterns, identified by the measurement outcomes $(1,1,1,0,0,0,0,0,0)$, $(0,0,0,1,1,1,0,0,0)$, and $ (0,0,0,0,0,0,1,1,1)$, are uniquely associated to the state $\ket{\Psi_{00}}$, and the success probability for the discrimination amounts to $P_s =1/81$.

\section{Methods}
\label{3}

\subsection{Analytical description of a BSM}
\label{3A}

    A BSM can be described by a set of POVM that accounts for the different statistics of light at the output, measured with photon-number-resolving detectors that are associated to certain input Bell states.
    The photon counts can be efficiently described via the formalism of generating functions.
    In particular, we can define the operator \cite{sperling2014quantum, sperling2018quasistates,engelkemeier2020quantum}
    \begin{equation}
        \label{generating}
        \hat{E}(x) = :\exp{\{-(1-x)\hat{n}\}}:,
    \end{equation}
    where $\hat{n}$ is the number operator counting the photons in the (here, single) mode under study.
    It is easy to see that the normal ordering---indicated via ``${:}\cdots{:}$''---in the definition of the generating functional is necessary to recover the photon number projector;
    i.e., it can be shown that
    \begin{equation}
        \ket{n}\bra{n} = \frac{1}{n!}\partial^{n}_x\hat{E}(x)|_{x=0},
    \end{equation}
    and this projector also describes the POVM element associated to a single-mode measurement of $n$ photons.

    Since the photon-counting statistics for a given input light state is obtained by calculating the Hilbert-Schmidt product among the density matrix $\hat{\rho}$ of the state and the given POVM element $\hat{\Pi}$, we can see how, under any operation $\hat{U}$ involving the input states, we can write
    \begin{equation}
    \begin{split}
        \Tr{\{\hat{\rho}'\hat{\Pi}\}}
        =&
        \Tr{\{\hat{U}\hat{\rho}\hat{U}^{\dagger}\hat{\Pi}\}}
        \\
        =& \Tr{\{\hat{\rho}\hat{U}^{\dagger}\hat{\Pi}\hat{U}\}}
        = \Tr{\{\hat{\rho}\hat{\Pi}'\}},
    \end{split}
    \end{equation}
    with
    \begin{equation}
        \label{nuovoprojector}
        \hat{\Pi}' = \hat{U}^{\dagger}\hat{\Pi}\hat{U}.
    \end{equation}
    
    Thus, by duality, the detection scheme evolves with the adjoint map of the operation $\hat{U}$.
    Therefore, we can describe any BSM based on photon-counting schemes by a set of operators of the type in Eq. \eqref{nuovoprojector}, and different operations $\hat{U}$ yield different efficiencies in the detection strategy.
    The aforementioned efficiency is again quantified by the success probability, which is now given by the unique probabilities computed with $d^2$ Hilbert-Schmidt products among the input Bell states $\rho_{ij}, \; i,j = 0,\dots,d-1$ defining the density matrix associated to the states defined in Eq. \eqref{bell} and the POVM set.
    These scalar products give counting statistics of the input states, and values of these distributions that are unique contribute towards in the counting of the averaged sum for the success probability.
    Therefore, in this formalism, we put all the evolution on the BSM since it is more convenient to describe the whole measurement apparatus by a POVM assigning to different input states a probability distribution without considering every state separately.

    In our specific scenario, including the appropriate multi-mode generalization, the set of POVM is described by the following operators
    \begin{equation}
        \label{initform}
        \hat{\Pi}' = \bigotimes_{k=0}^{2d-1} \frac{1}{n_k!}\partial^{n_k}_{x_{k}}\hat{U}^{\dagger}_{BS}\hat{S}^{\dagger}\hat{E}(x_k)\hat{S}\hat{U}_{BS}|_{x_{k}=0},
    \end{equation}
    with $U_{BS}$ defining the beam splitter transformation on the creation operators $\hat{a}^{\dagger}_{i}, \; i=0 \dots 2d-1$ of photons in the different paths (i.e., spatial modes) as
    \begin{equation}
        \hat U_{BS}\hat{a}^{\dagger}_{i}\hat U_{BS}^{\dagger} = \sum_{k}U_{ik}\hat{a}^{\dagger}_{k}
    \end{equation}
    where $U_{ik}$ are the matrix elements of the following unitary transformation
    \begin{equation}
        \label{BS}
        \left( U_{ik}\right)_{i,k\in\{0,\dots ,2d-1\}}
        =\frac{1}{\sqrt{2}}
        \begin{pmatrix}
            1 & 0  & \dots & 0 & i & 0 & \dots & 0\\
            
            0 & 1 &\dots & 0 & 0 & i & \dots &0\\
            0 & 0 & \dots & 1 & 0 & 0 & \dots & i\\
            \vdots & \vdots & \ddots & \vdots & \vdots & \vdots & \ddots & \vdots \\
            i & 0 & \dots & 0 & 1 & 0 &\dots & 0\\
            0 & i & \dots & 0 & 0 & 1 &\dots & 0 \\
            0 & 0 & \dots & i & 0 & 0 &\dots & 1 \\
        \end{pmatrix},
    \end{equation}
    as well as the above $\hat{S}$ the squeezing operator.
    This means that each element of the POVM set is characterized by the number of photons detected and the value of the squeezing we choose.

    Squeezing is one of the building blocks of continuous-variable quantum information processing \cite{braunstein2005squeezing} as it is utilized to build up Gaussian states beyond non-squeezed coherent states.
    Squeezing approximates one of the lowest-order light-matter interaction, i.e., how electromagnetic fields change in presence of a non-linear medium.
    Namely, the interaction Hamiltonian for the process is quadratic in the creation and annihilation operator, rendering the dynamics of a given state easy to compute.
    One can define the squeezing operator as \cite{vogel2006quantum}
    \begin{equation}
        \hat{S}(z) = \exp\left(\frac{1}{2}z^\ast\hat{a}^2-\frac{1}{2}z\hat{a}^{\dagger 2}\right),
    \end{equation}
    where $z = r e^{i\phi}$ is dubbed the squeezing parameter, a complex number related to the strength of non-linearity of the medium.
    The action on a creation operator is given by
    \begin{equation}
        \hat{S}(z)\hat{a}^{\dagger}\hat{S}(z)^{\dagger} = \mu \hat{a}^{\dagger}+\nu^\ast\hat{a}.
    \end{equation}
    In this Bogolubov transformation, the mixing of $\hat{a}$ and $\hat{a}^{\dagger}$ operators for the same mode occurs.
    The unitarity condition imposes the conservation of canonical equal-time commutators and yields the relation
    \begin{equation}
        \mu^2 - |\nu|^2 = 1,
    \end{equation}
    allowing one to identify the parameters $\mu$ and $\nu$ as
    \begin{equation}
        \label{relmunu}
        \mu = \cosh{(r)}
        \quad\text{and}\quad
        \nu = e^{i\phi}\sinh{(r)}.
    \end{equation}
    The squeezing operator can also be parametrized in terms of the ratio
    \begin{equation}
        \frac{\nu}{2\mu} = \frac{1}{2}e^{i\phi}\tanh{(r)}
        \stackrel{\text{def.}}{=}\delta.
    \end{equation}
    Also, we can relate the values of the $r$ parameter with the usual unit measurement of squeezing $\zeta$ in decibel (dB) as
    \begin{equation}
        \zeta\,(\mathrm{dB}) = -10\log_{10}{e^{-2r}}.
    \end{equation}

    In Ref. \cite{zaidi2013beating}, it was shown how the number of unique click patterns detected by a Bell measurement can be enhanced by placing squeezers before a photon-number-resolving detector.
    Also, squeezing is a parity preserving operation, meaning that an odd or even number of photons in a given mode remains odd or even, respectively.
    Since the click pattern detection for the usual BSM of qubits is based on the parity of the counted photons in each path, Bell states that can be discriminated without squeezing remain distinguishable.
    In addition, however, it can be seen that this approach partially removes the degeneracy of the $\ket{\Phi^{\pm}}$ states for a given interval of values of the squeezing parameter, and it ideally boosts the success probability up to $P_s= 64.3\%$.
    Squeezing the input photon-number states has the following effects.
    Firstly, squeezing a Bell state changes the amplitudes of any state in the superposition describing it.
    In particular, the amplitudes acquire a dependency on $r$ that goes as a combination of hyperbolic functions as given in Eq. \eqref{formulasqueezing}.
    (Such characteristic profiles of coefficients are visible in the plot of the success probability of Bell measurements.)
    This alteration of amplitudes, in turn, can cancel some of the amplitudes related to photon-number states, meaning that squeezing states with the same parity can produce an additional degree of interference on top of the one given by the mixing of modes on a beam splitter.

\subsection{Numerical simulation benchmark}
\label{3B}

    To carry out the numerical simulation of the dynamics given by our BSM, it is convenient to rely on a normal-ordered form of the squeezing operator \cite{vogel2006quantum}
    \begin{equation}
    \label{espansionesqueezing}
        \hat{S}(z)
        =
        \exp\left(-\delta\hat{a}^{\dagger 2}\right)
        \left(\frac{1}{\mu}\right)^{\hat{n}+\frac{1}{2}}\exp\left(\delta\hat{a}^2\right),
    \end{equation}
    assuming $\delta$ is a real number.
    With the relation \eqref{espansionesqueezing}, we can calculate the action of the squeezing operator on an $n$-particle photon-number state \cite{nieto1997displaced} as
    \begin{equation}
        \label{formulasqueezing}
    \begin{split}
        &\hat{S}(r)\ket{n}
        \\
        =&
        \frac{\sqrt{n!}}{\cosh{(r)}^{n+\frac{1}{2}}}\sum_{j=0}^{\lceil \frac{n}{2} \rceil}
        \frac{\Big(-\frac{\tanh{(r)}}{2}\Big)^j\cosh{(r)}^{2j}}{(n-2j)!j!}
        \\
        &\times
        \sum_{k=0}^{\infty}\frac{\Big(\frac{\tanh{(r)}}{2}\Big)^k\sqrt{(n{-}2j{+}2k)!}}{k!}\ket{n-2j+2k}.
    \end{split}
    \end{equation}

\begin{figure}
    \centering
    \includegraphics[width=0.5\textwidth]{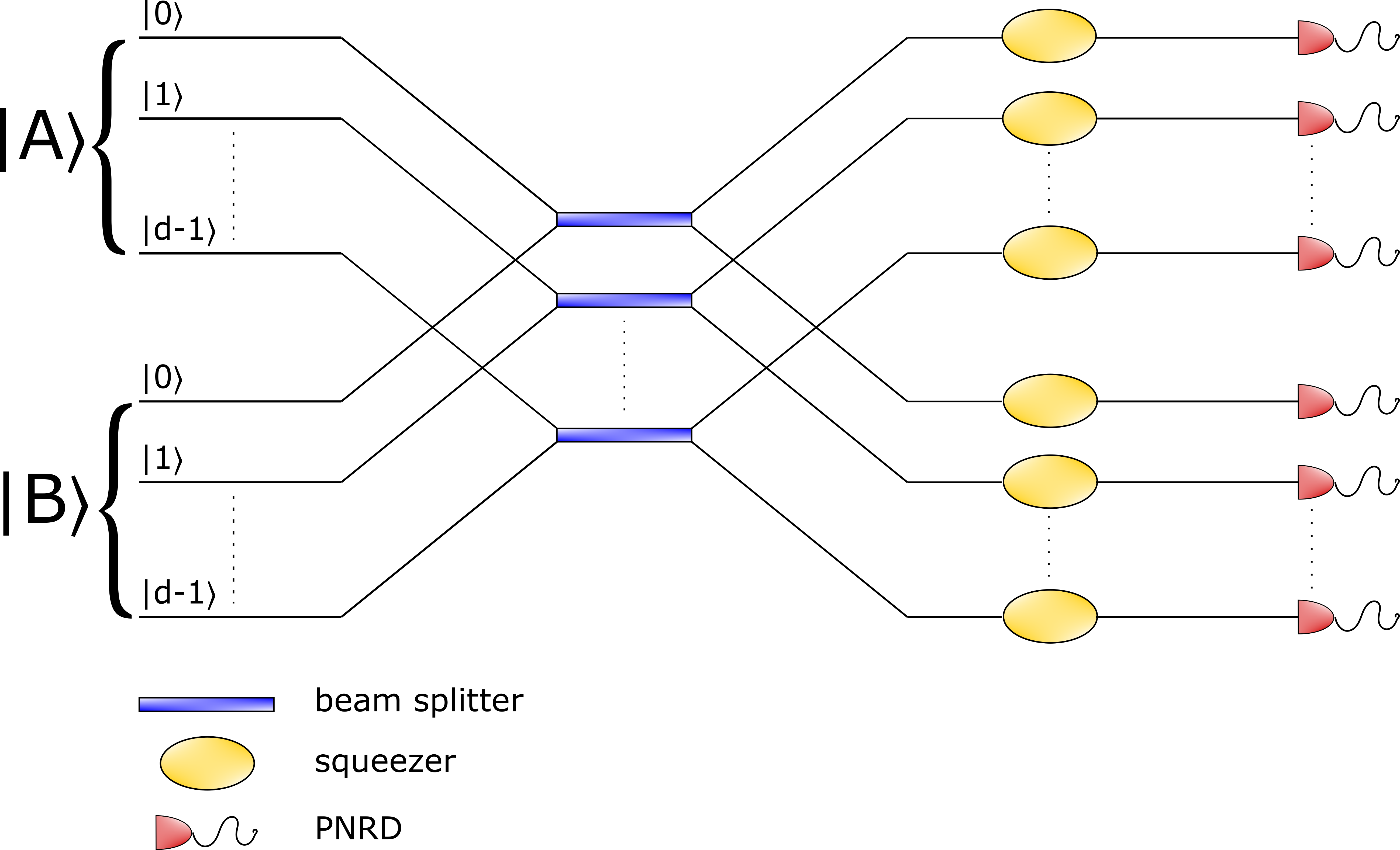}        
    \caption{%
            The pre-detection squeezing scheme for a two-qudit system.
            Paths in the figure corresponds to logical levels of the qudits.
            These levels are mixed via $d$ beam splitters, and the output are squeezed with an equal intensity across the modes before the photon-number measurement.
            }\label{qutrit}
\end{figure}

\begin{figure}
    \centering
    \includegraphics[width=0.5\textwidth]{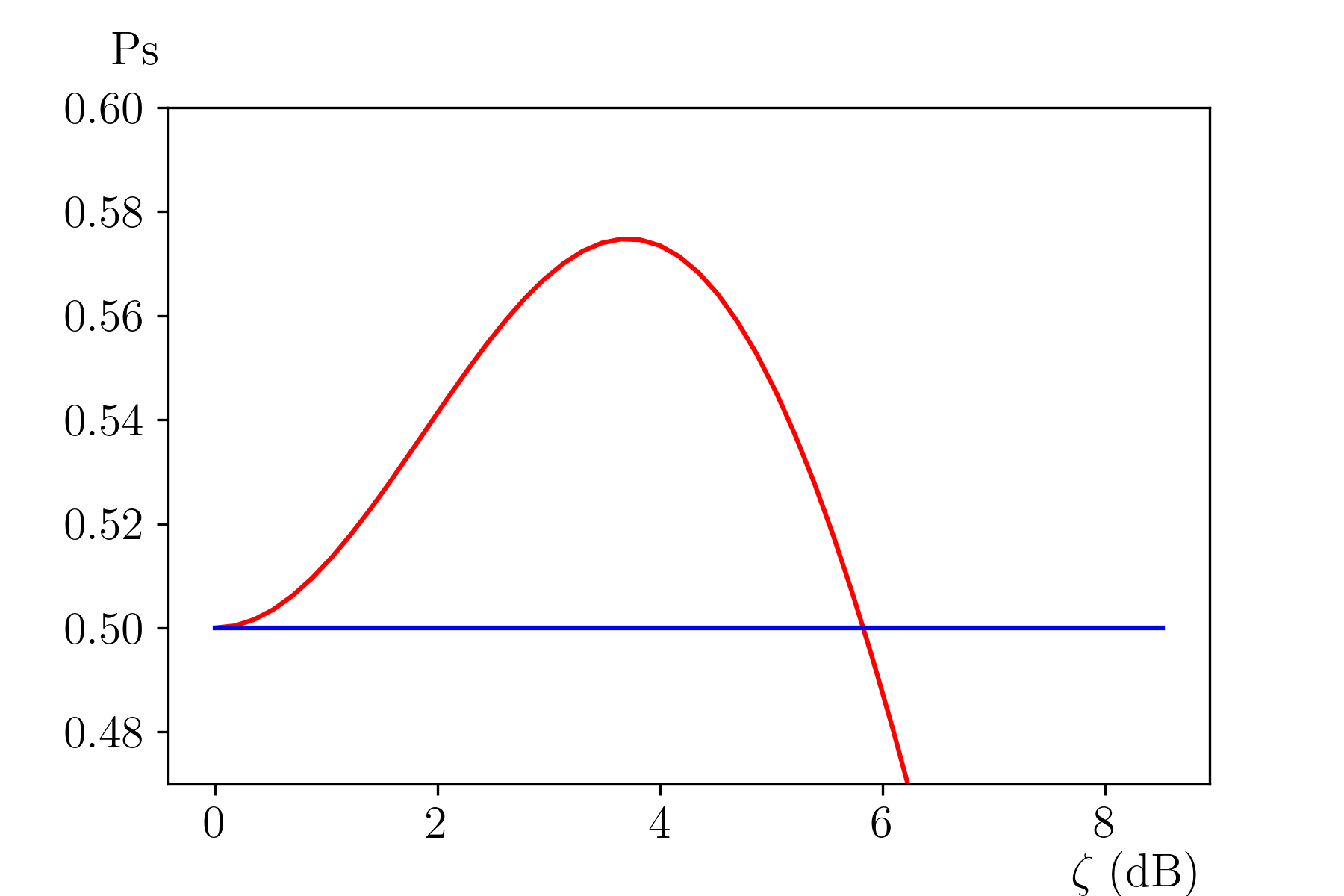}
    \caption{%
        Boosted probability for a two dimensional BSM with pre-detection squeezing (red, dashed curve).
        For a limited photon-number resolution of five, there is a range of values $r \in (0,1]$ that provides an advantage over the linear optical Bell discrimination (blue, solid line), with a peak at $P_s \approx 0.5747$.
        }\label{qubitss}
\end{figure}

    As a simple test, we reconstruct the success probability as a function of a squeezing parameter for the two-qubit case.
    The parameters are taken to be real ($\phi_k = 0, \; \forall k = 0 \dots d-1$) and uniform ($r_k = r, \; \forall k = 0 \dots d-1$), and we also assume the photon-number resolution of the detectors to be upper bounded by five photons.
    This means that the summation over the index $k$ in Eq. \eqref{formulasqueezing} should be truncated to an $k_{\max} = \lfloor \tfrac{n_{\max}-n+2j}{2} \rfloor$ with $n_{\max} = 5$ being the assumed resolution.
    In Fig. \ref{qubitss}, we plot the squeezing parameter versus the overall success probability of discriminating two dimensional Bell states, $P_s$.
    The plot is calculated using 50 points of the squeezing parameter, spanning the interval $(0,1]$ with a step size of $r = 0.02$.
    A boost of the success probability up to $P_s \approx 0.575$ for $r \approx 0.44$ can be observed, further corresponding to $\zeta \approx 3.82\,\mathrm{dB}$ and which is in agreement with Fig. 4 of Ref. \cite{kilmer2019boosting}.
    Beyond this benchmark for the two-dimensional case, a comprehensive analysis for the photon-number resolution and losses can be found in Ref. \cite{kilmer2019boosting}.

\section{Results}
\label{4}

\subsection{Analytical description of the pre-detection squeezing POVM}
\label{4A}

    Our implementation of the pre-detection squeezing scheme for high-dimensional BSM relies on $d$ beam-splitter coupling of pairwise paths of the uncorrelated incoming photons.
    Specifically, we mix the mode $k$ with the mode $k' = k\oplus d$, where ``$\oplus$'' denotes addition modulus $d$.
    A schematic representation can be found in Fig. \ref{qutrit}.
    After some algebra, provided in Appendix \ref{Appendix}, the closed form of our POVM can be put into the form
    \begin{equation}
        \label{formulafinale}
    \begin{split}
        \hat\Pi' =& \prod_{k=0}^{d-1}\partial^{n_k}_{x_k}\frac{1}{\sqrt{d(x_k)}}
        \\&
        \times
        :\exp\left\lbrace
            \frac{\lambda(x_k){-}\lambda(x_{k'})}{2}
            \left(
                \hat{a}^{\dagger 2}_k
                {+}\hat{a}_k^2
                {-}\hat{a}^{\dagger 2}_{k'}
                {-}\hat{a}^2_{k'}
            \right)
        \right.
        \\
        &
        -i\left[\lambda(x_k) +\lambda(x_{k'})\right]
        \left(
            \hat{a}^{\dagger}_{k}\hat{a}^{\dagger}_{k'}-\hat{a}_{k}\hat{a}_{k'}
        \right)
        \\
        &
        +\frac{\theta(x_k)+\theta(x_{k'})}{2}
        \left(
            \hat{a}^{\dagger}_{k}\hat{a}_k+\hat{a}^{\dagger}_{k'}\hat{a}_{k'}
        \right)
        \\
        &
        \left.
            +i\frac{\theta(x_k)-\theta(x_{k'})}{2}
            \left(
                \hat{a}^{\dagger}_k\hat{a}_{k'}-\hat{a}^{\dagger}_{k'}\hat{a}_k
            \right)
        \right\rbrace:.
    \end{split}
    \end{equation}
    In principle, each mode comes with its own parameter $r_k$---such dependences are encompassed in the $\lambda(x_k)$ and $\theta(x_k)$ coefficients; cf. Appendix \ref{Appendix}.
    The optimality of the POVM set can thus be found by maximizing the success probability as a function of the squeezing.

\subsection{Numerical simulations for the high-dimensional case}
\label{4B}

\begin{figure}
    \centering
    \includegraphics[width=0.5\textwidth]{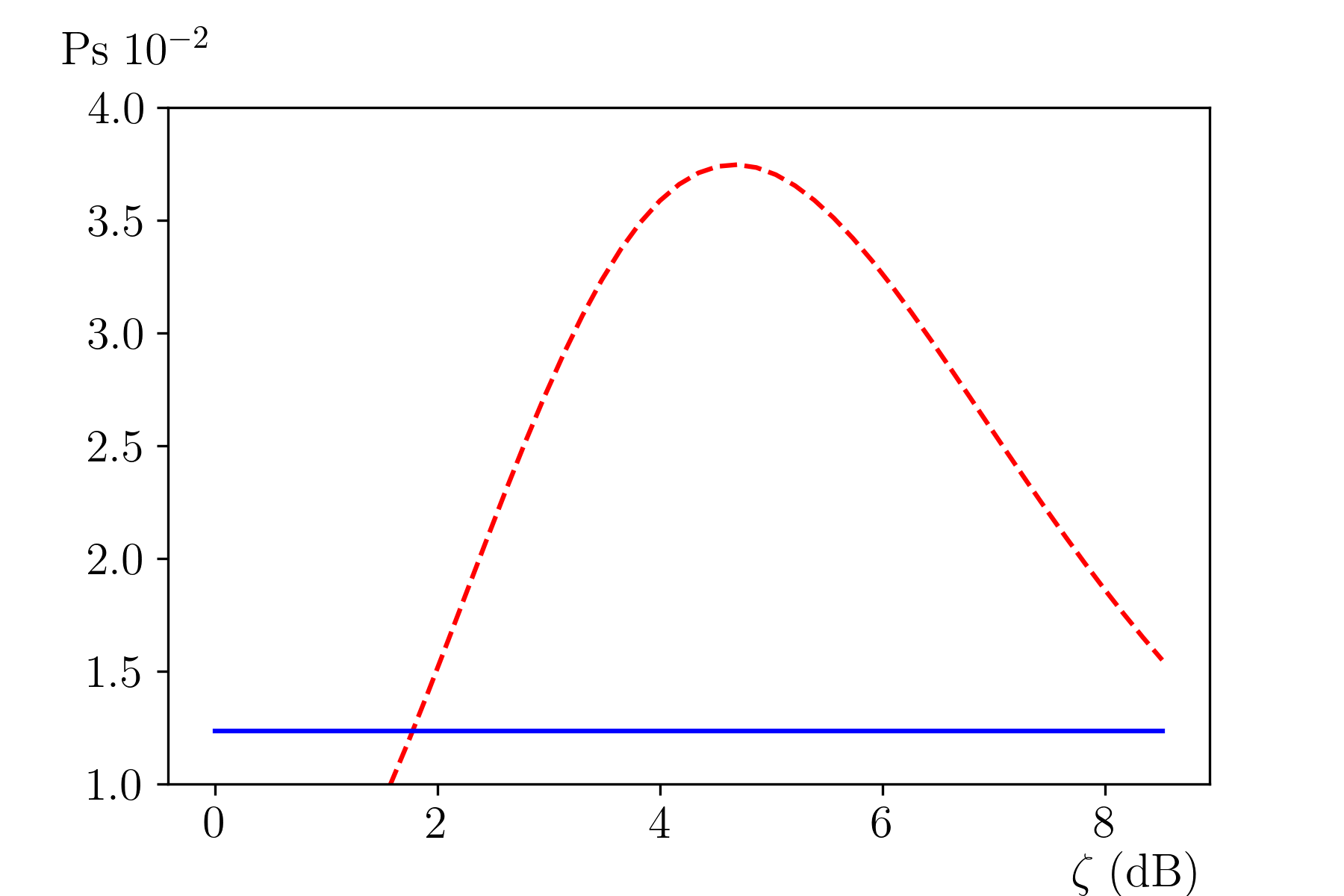}
    \caption{%
        Boosted probability for a three dimensional BSM with pre-detection squeezing (red, dashed curve).
        Again, with a maximal photon resolution of five, we see that, after a value corresponding to $r \approx 0.24$ ($\zeta \approx 2.08\,\mathrm{dB}$), we always beat the limit in the success probability given by linear optics plus auxiliary photon (blue, solid line).
        The peak in the Bell-state distinguishability has a value of $P_s \approx 0.0374$ and happens at $r \approx 0.56$ ($\zeta \approx 4.86\,\mathrm{dB}$).
    }\label{trist}
\end{figure}

    We can now simulate the pre-detection squeezing scheme to qudit systems.
    By relying on the same conditions as in the two-dimensional simulation, we plot in Fig. \ref{trist} the success probability given by our set of POVM in Eq. \eqref{formulafinale} with a limited photon-number resolution of five photons.
    We compare this finding with the $P_s=1/81$ limit obtained with linear optics and auxiliary modes in Ref. \cite{luo2019quantum}.
    We see that the squeezed probability beats the linear optical one after $r \approx 0.24$, corresponding to a very moderate value of $\zeta \approx 2.08\,\mathrm{dB}$.
    Furthermore, a peak in the discrimination $P_s \approx 0.0374$ is obtained for $r \approx 0.56$, viz. $\zeta \approx 4.86\,\mathrm{dB}$.
    The clear advantage in the success probability is further strengthened by the absence of an auxiliary single photon and of a QFT\textsubscript{3} transformation on the photon space, which is experimentally demanding. This result is not trivial because, for two-dimensional systems, the squeezing procedure does not beat even the most simple BSM with auxiliary photons;
    see also Refs \cite{griceancilla, ewertancilla}.

\begin{figure}
    \centering
    \includegraphics[width=0.5\textwidth]{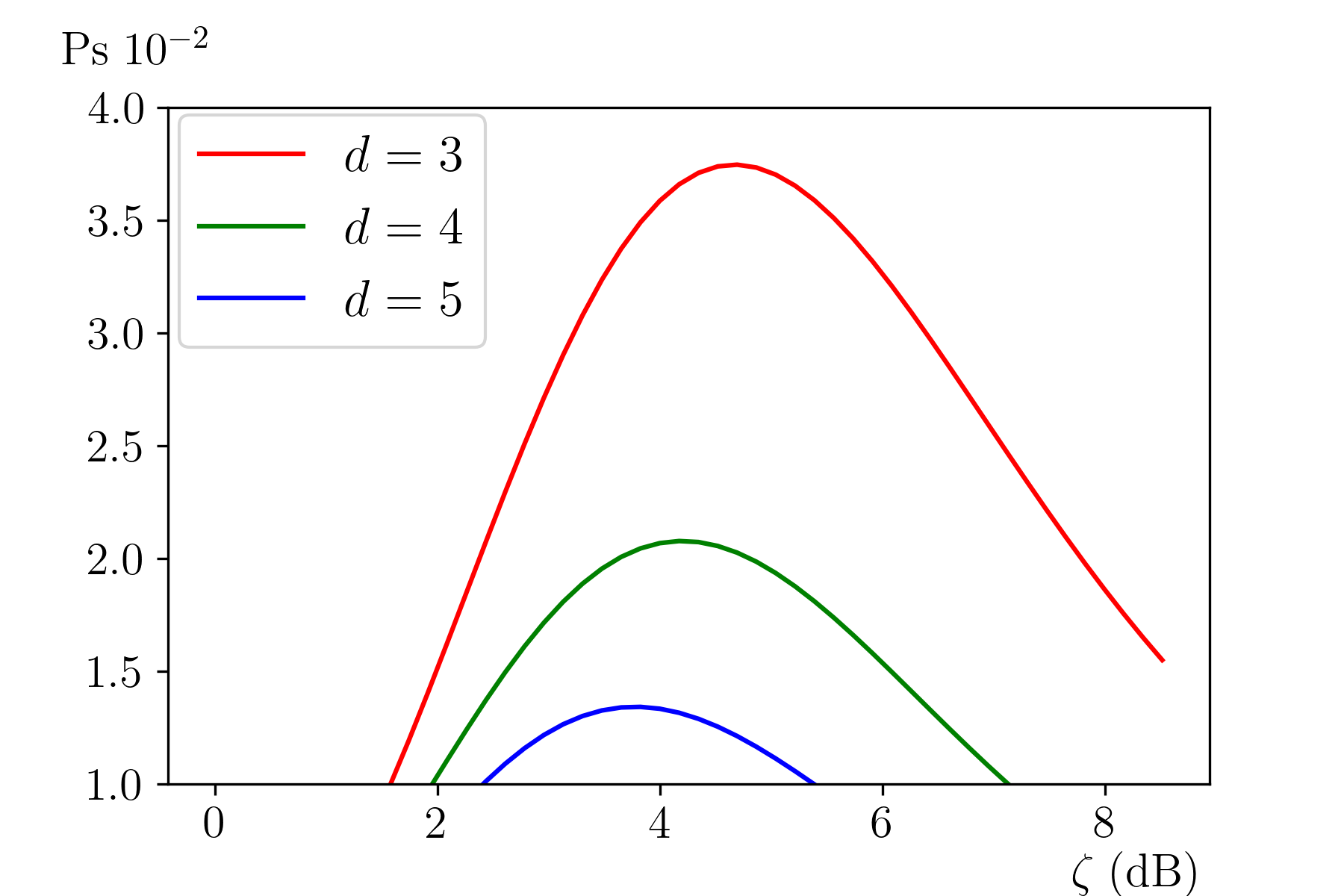}
    \caption{%
        Boosted probability for different dimensions of BSMs with pre-detection squeezing .
    }\label{quart}
\end{figure}

    We also simulated cases for the higher dimensions $d=4$ and $d=5$, obtaining further boosting as clearly shown in Fig. \ref{quart}
    For the $d=4$ case, the peak of $P_s \approx 0.0208$ is obtained for $r \approx 0.5$ ($ \zeta \approx 4.34\,\mathrm{dB}$), and we find $P_s \approx 0.0134$ at $r \approx 0.46$ ($\zeta \approx 4\,\mathrm{dB}$) for $d=5$.
    Again, neither auxiliary photons nor high-dimensional QFT are required to remove the degeneracy among the Bell states;
    we just need to add more beam splitters, linking the modes pairwise. Therefore, this approach greatly simplify the structure of the BSM, together with advantageous scalability.
    The boost in these high-dimensional probabilities is mainly given by the uniqueness of the vacuum state, and the non-degenerate clicks are completely associated with the $\ket{\Psi_{00}}$ state.
    The list of unique click patterns corresponding to the peaks in $P_s$, together with their probability amplitudes, is shown in Table \ref{tab:multirow-color}.

\begin{table}
    \centering
    \caption{%
        Most probable photon-number measurement patterns (shorthand ``click pattern'') together with their amplitudes for different dimensions
        Note that the vacuum state is always the dominant term in the sum of the squared unique amplitudes that determines the success probability.
        These click patterns all are uniquely associated with the $\ket{\Psi_{00}}$ state.
    }\label{tab:multirow-color}
    \begin{tabular}{ccc}
        \hline\hline
        dimension & amplitude & click pattern \\
        \hline
        \multirow{3}{*}{$d=3$} 
        & $-0.5623i$ & (0,0,0,0,0,0)  \\ \cline{2-3}
        & \multirow{3}{*}{$0.0489i$} & (0, 0, 0, 3, 2, 1)\\
        &  & (0, 0, 4, 0, 3, 1) \\
        &  & (0, 4, 1, 1, 3, 1) \\
        &  & (0, 4, 4, 3, 4, 1) \\
        &  & (4, 4, 4, 2, 3, 1)\\
        &  & (5, 0, 2, 4, 4, 1)\\
        &  & (5, 4, 0, 0, 4, 1) \\
        &  & (5, 4, 3, 3, 0, 1)\\
        \hline
        \multirow{3}{*}{$d=4$} & $-0.5716i$ & (0, 0, 0, 0, 0, 0, 0, 0) \\ \cline{2-3}
        & \multirow{3}{*}{$0.0175i$} & (0, 0, 0, 0, 4, 0, 3, 3) \\ & & (0, 0, 0, 4, 4, 4, 1, 3)\\
        & & (0, 1, 0, 0, 3, 2, 1, 3)\\ 
        & & (0, 1, 0, 4, 4, 0, 4, 3)\\
        & & (1, 1, 0, 0, 0, 0, 1, 3)\\
        & & (1, 1, 0, 4, 0, 3, 4, 3)\\
        & & (1, 1, 4, 4, 4, 1, 4, 3)\\
        & & (1, 2, 0, 4, 0, 0, 2, 3)\\
        \hline
        \multirow{3}{*}{$d=5$} & $-0.5767i$ & (0, 0, 0, 0, 0, 0, 0, 0, 0, 0) \\ \cline{2-3}
        & \multirow{3}{*}{$0.0051i$} & (0, 0, 0, 0, 0, 4, 4, 4, 2, 0) \\ 
        & & (0, 0, 0, 1, 0, 4, 4, 1, 2, 0) \\
        & & (0, 0, 1, 1, 0, 4, 1, 1, 2, 0) \\
        & & (0, 0, 1, 2, 0, 4, 0, 3, 2, 0) \\
        & & (0, 1, 2, 1, 0, 0, 3, 1, 2, 0) \\
        & & (0, 1, 2, 2, 0, 0, 2, 3, 2, 0) \\
        & & (0, 1, 3, 1, 4, 4, 4, 3, 2, 0) \\
        & & (0, 1, 3, 2, 4, 4, 4, 0, 2, 0) \\
        & & (1, 3, 3, 0, 0, 4, 0, 1, 2, 0) \\
        & & (1, 3, 3, 1, 0, 3, 4, 3, 2, 0) \\
        & & (1, 3, 4, 1, 0, 3, 1, 3, 2, 0) \\
        & & (1, 3, 4, 2, 0, 3, 1, 0, 2, 0) \\
        & & (2, 0, 0, 0, 4, 4, 3, 3, 2, 0) \\
        & & (2, 0, 0, 1, 4, 4, 3, 0, 2, 0) \\
        & & (2, 0, 1, 1, 4, 4, 0, 0, 2, 0) \\
        & & (2, 0, 1, 2, 4, 3, 4, 2, 2, 0) \\
        \hline\hline
    \end{tabular}
\end{table}

    To strengthen our results, we also tested different scenarios for the squeezers.
    For example, we noticed that a finer grained scan over the squeezing parameter interval yields no significant details on the maximum, while the choice of different resolutions has an impact only in the two-dimensional case and not in the high-dimensional one.
    Furthermore, relaxing the uniformity constraint on the squeezing parameters by allowing for different intensities does not result in any improvement to the success probability, and the same statement is true if we allow the squeezing to be non-uniform and complex (i.e., $\phi_k \neq 0 \; \forall k=0,\dots d-1$).
    Finally, we can also report that combining the ancilla-aided linear optical setup discussed in Sec. \ref{2B} with the pre-detection scheme does not bring any benefit to the success probability.
    We interpret this fact as if no interference concurs to the uniqueness of click patterns, that are fixed by the structure of the setup, while the decreasing behavior in the large-squeezing region of the amplitude presented in Eq. \eqref{espansionesqueezing} holds true, reflecting into a monotonically decreasing success probability.

\section{Discussion}
\label{5}
    It is well understood that nonlinearities are required to discriminate high-dimensional Bell states, which are fundamental in quantum information and communication.
    We find that squeezing proves to be more efficient than resorting to the use of auxiliary single photon quantum states, whose implementation is often experimentally challenging. 
    Furthermore, our method is scalable from a theoretical point of view since, by increasing the dimensionality of the Hilbert space, the success probability of BSM is higher as compared to the case when ancillary photons are considered.
    It is also worth pointing out that, to realize an experimental apparatus performing the Bell measurement, we do not need to create $d-2$ auxiliary photons and a $d$-dimensional quantum Fourier transform;
    the requirement of our setup is to superimpose modes in $d$ beam splitters and then squeeze their joint output.
    The squeezing values needed to obtain the maximum success probability are practical and reachable with the currently available setups \cite{baboux2023nonlinear}.

\begin{figure}
    \centering
    \includegraphics[width=0.5\textwidth]{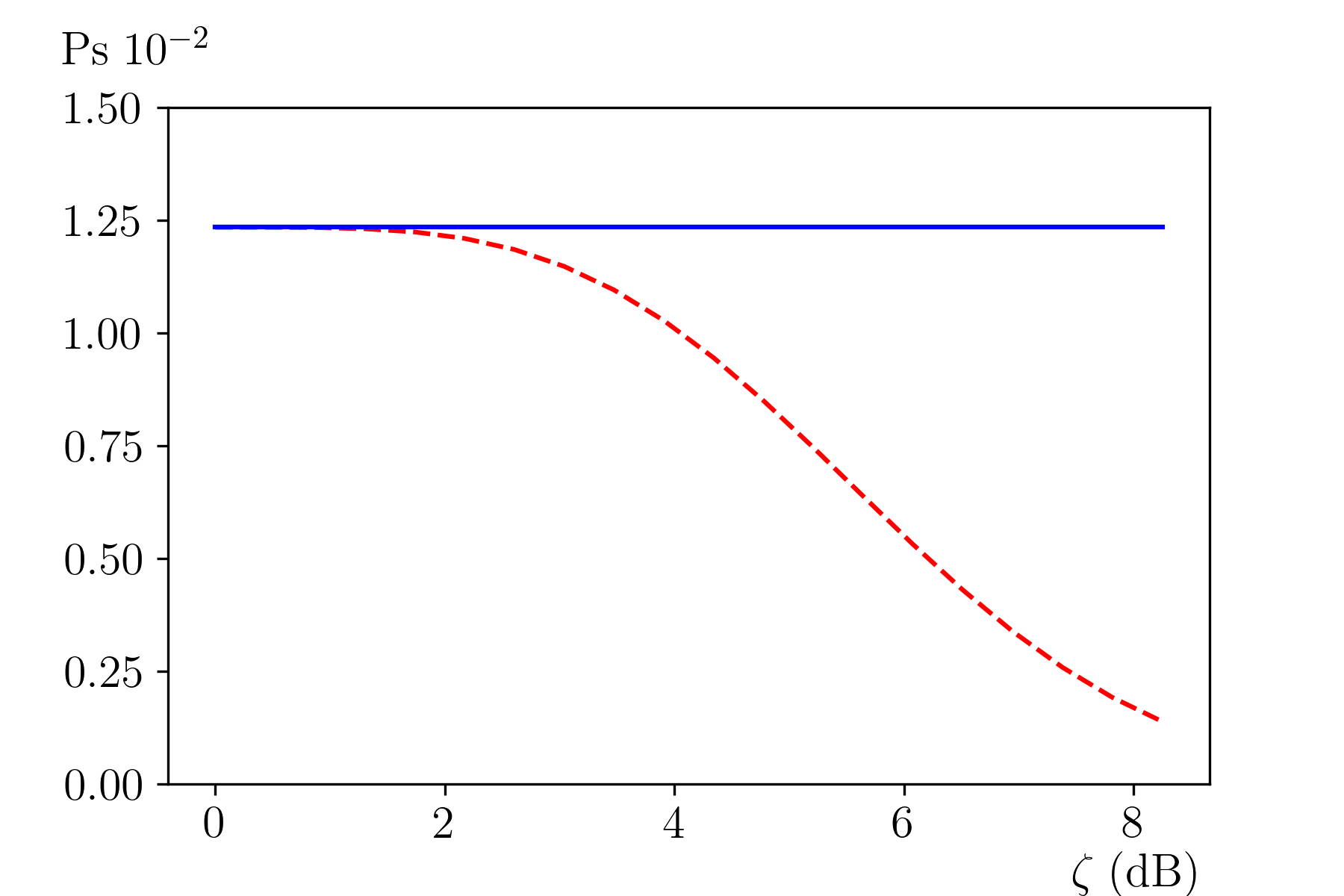}
    \caption{%
        Comparison between the auxiliary-single-photon-aided linear optical setup (blue, solid line) and the same setup but with squeezers before photodetection (red, dashed curve). The numerical simulation shows that combining squeezing and ancillary photons does not bring any benefit to the success probability for Bell state discrimination.
    }
    \label{combining}
\end{figure}

    Further improvements can consider the study of the analytical description for the POVM of our Bell-state analyzer by introducing dark counts and photon losses in the description of the generating function. This would further advance our description to a more accurate and physically meaningful level.
    Moreover, the analytical approach points a way to discover the possibilities of nonlinear optics for Bell-state discrimination;
    i.e., having a closed, exact form for the POVM can be useful in order to propose new criteria in the analysis of Bell measurements.
    This is also suggested by Ref. \cite{calsamiglia2002generalized}, where the no-go theorem for linear optical Bell measurements is proven by finding the POVM elements associated to input Bell states.
    Therefore, the joint description of the click-patterns provided by the generating function approach can be useful to find scaling laws and bounds for the efficiency of different nonlinear setups, as functions of the dimension of the quantum system.
    Interestingly, it was noticed that the pre-detection squeezing method is not compatible with the auxiliary-single-photon-aided setup proposed for linear optics, as can be seen in Fig. \ref{combining}.
    This poses new questions about whether a given set of unitaries is compatible with high-dimensional Bell measurements or not.
    On the other hand, the realization of this BSM would impose new challenges from an experimental point of view since one has to deal with additional effects such as extra loss channels, squeezing imperfections and excess noise after the squeezing procedure, likely having an impact on the success probability.

\section{Conclusion}
\label{6}

    In this work, we investigated the possibility of realizing a high-dimensional Bell state analyzer supported by squeezer devices.
    We provided an analytical description of our setup in terms of the POVM that describes the squeezed photon-number statistics, and we simulated the success probability of incomplete Bell measurements for different dimensions.
    Our results demonstrated a positive outcome, overcoming previous theoretical proposals for dimensions 3 and 4 showing the validity of this method. In addition, our proposal is scalable in terms of dimensionality and experimentally implementable, thus opening a new path for qudits-based quantum networks.

    
    

\begin{acknowledgments}
    L.B thanks Riccardo Cioli for useful comments and discussions.
    This research has been cofunded by the European Union ERC StG, QOMUNE, 101077917, and by the NextGeneration EU, "Integrated infrastructure initiative in Photonic and Quantum Sciences" - I-PHOQS [IR0000016, ID D2B8D520, CUP B53C22001750006].
    J.S. acknowledges funding through the Deutsche Forschungsgemeinschaft (DFG, German Research Foundation) via the transregional collaborative research center TRR 142 (Project C10, Grant No. 231447078).
\end{acknowledgments}

\appendix

\section{Pre-detection squeezing BSM analytical description}
\label{Appendix}

    To evaluate the action of the squeezing and the beam-splitter operators on the POVM, we need the following formulae.
    First of all, it holds true that
    \begin{equation}
    \begin{split}
        \label{squeezingaction}
        &\hat{S}(z) \ket{\alpha}= \ket{z, \alpha}
        \\
        =&
        \frac{1}{\sqrt{\mu}}\exp\left\lbrace
            -\frac{1}{2}|\mu\alpha-\nu\alpha^\ast|^2+(\mu\alpha-\nu\alpha^\ast)\hat{a}^{\dagger}
        \right.
        \\
        &\phantom{\frac{1}{\sqrt{\mu}}\exp}\left.
            -\frac{\nu}{2\mu}[\hat{a}^{\dagger}+(\nu\alpha-\mu\alpha^\ast)]^2
        \right\rbrace\ket{0},
    \end{split}
    \end{equation}
    for $\ket{\alpha}$ a coherent state of light.
    The state $\ket{r, \alpha}$ is referred to as a squeezed coherent state, and it is clear how it can be cast in terms of creation operators.
    Furthermore, for coherent states, the completeness relation reads
    \begin{equation}
        \label{compl}
        \hat{I} = \int\frac{d^2\alpha}{\pi}\ket{\alpha}\bra{\alpha},
    \end{equation}
    where the integration is performed over the whole complex plane.
    Another useful formula is the overlap between two coherent states,
    \begin{equation}
        \label{ON}
        \braket{\beta}{\alpha} = \exp\left\{\frac{1}{2}
            \left(-|\beta|^2-|\alpha|^2+\beta^\ast\alpha\right)
        \right\}.
    \end{equation}
    For complex numbers, a Gaussian integration can be performed by means of the following formula:
    \begin{equation}
        \label{gaussian}
    \begin{split}
        &\int d\xi^2\exp\left\{-z|\xi|^2-x\xi^2-y\xi^{*2}+u\xi+v\xi^\ast\right\}
        \\
        =&\frac{\pi}{\sqrt{z^2-4xy}}\exp\left\{\frac{zuv-yu^2-xv^2}{z^2-4xy}\right\},
    \end{split}
    \end{equation}
    with the positive branch of the square-root defined for $\mathrm{Re}(z^2-4xy)>0$.
    This results still holds for an expression that is in normal ordering since, under this prescription, the operators close a commuting algebra and can be treated as complex numbers.

    Now, we are ready to evaluate the expression in Eq. \eqref{initform}.
    Firstly, we can restrict our attention to a single mode $k$ undergoing a squeezing transformation.
    We thus omit the subscript $k$ here, while keeping in mind how the full result will be just the tensor product of $2d$ copies of the operator below.
    By using twice the identity \eqref{compl}, we obtain
    \begin{equation}
        \label{pipo}
    \begin{split}
        &\hat{P}(x)
        \stackrel{\text{def.}}{=}
        \hat{S}^{\dagger}(r)\hat{\Pi}\hat{S}(r)
        = \hat{S}^{\dagger}(r)\hat{I}\hat{\Pi}\hat{I}\hat{S}(r)
        \\
        =&\int\frac{d^2\alpha}{\pi}\int\frac{d^2\beta}{\pi}
        \hat{S}^{\dagger}(r)\ket{\alpha}\braket{\alpha}{\hat{E}(x)|\beta}\bra{\beta}\hat{S}(r).
        \end{split}
    \end{equation}
    Since $\hat{E}(x)$ is a normally ordered operator, the operators $\hat{a}, \ \hat{a}^{\dagger}$ can be directly substituted with their eigenvalues corresponding to the coherent states $\ket{\alpha}, \ \ket{\beta}$,
    \begin{equation}
        \braket{\alpha}{\hat{E}(x)|\beta}
        =
        \exp\left\{
            x\alpha^\ast\beta-\frac{|\beta|^2+|\alpha|^2}{2}
        \right\}.
    \end{equation}
    By using Eq. \eqref{squeezingaction}, we can see that
    \begin{equation}
        \hat{S}^{\dagger}(r)\ket{\alpha} = \ket{-r, \alpha}
        \quad\text{and}\quad
        \bra{\beta}\hat{S}(r) = \bra{-r, \beta}
    \end{equation}
    hold true.
    From the relation \eqref{relmunu}, we have $\mu(-r) = \mu(r)$ and $\nu(-r)=-\nu(r)$, and the definition of the generating function for the photon-number projectors in Eq. \eqref{generating} implies $\ket{0}\bra{0} = :\exp{\{-\hat{a}^{\dagger}\hat{a}\}}:$.
    Equation (\ref{pipo}) then becomes
    \begin{equation}
        \begin{split}
            \hat P(x)=&
            \int\frac{d^2\alpha}{\pi}\int\frac{d^2\beta}{\pi}\frac{1}{\mu}
            \exp\left\lbrace
                -\frac{1}{2}|\mu\alpha+\nu\alpha^\ast|^2
            \right.
            \\
            &\left.
                +(\mu\alpha+\nu\alpha^*)\hat{a}^{\dagger}
                +\frac{\nu}{2\mu}[\hat{a}^{\dagger}-(\nu\alpha+\mu\alpha^*)]^2
            \right\rbrace
            \\
            &\times:\exp\left\{
                -\hat{a}^{\dagger}\hat{a}
                +x\alpha^\ast\beta
                -\frac{|\alpha|^2}{2}
                -\frac{|\beta|^2}{2}
            \right\}{:}
            \\
            &\times
            \exp\left\{
                -\frac{1}{2}|\mu\beta+\nu\beta^\ast|^2
                +(\mu\beta^\ast+\nu\beta)\hat{a}
            \right.
            \\
            &\left.
                +\frac{\nu}{2\mu}[\hat{a}-(\nu\beta^*+\mu\beta)]^2
            \right\}.
        \end{split}
    \end{equation}
    which is clearly a normally ordered expression.
    This allows us to apply Eq. \eqref{gaussian} twice to resolve the complex Gaussian integrals.
    This finally yields
    \begin{equation}
        \hat{P}(x) = \frac{1}{\sqrt{d(x)}}
        :\exp\left\{
            \lambda(x)(\hat{a}^{\dagger 2}+\hat{a}^2)+\theta(x)\hat{a}^{\dagger}\hat{a}
        \right\}:,
    \end{equation}
    where
    \begin{equation}
    \begin{split}
        \lambda(x) =& \frac{\nu}{2\mu}\left[1-\frac{x^2}{d(x)}\right],
        \\
        \theta(x) =& -\left[1-\frac{x}{d(x)}\right],
        \\
         d(x) =& \mu^2-x^2\nu^2,
    \end{split}
    \end{equation}
    which correctly reproduces the squeezed vacuum projector in the limit of $x\rightarrow 0$ and returns the generating function in Eq. \eqref{generating} for $r\rightarrow 0$.

    The above operator now needs to be transformed through beam-splitter transformations.
    We remind ourselves that this transformation here connects the input modes pairwise, meaning that we should consider together the coupled modes $k$ and $k'=k\oplus_{d} d$ to consider the commutation relations among ladder operators of the same mode.
    This amounts to consider the transformed tensor product
    \begin{equation}
        \label{casino}
        \hat{U}^{\dagger}_{BS}\hat{P}_{k} (x)\otimes\hat{P}_{k'}(y)\hat{U}_{BS}.
    \end{equation}
    We further employ $\hat{I} = \hat{U}_{BS}\hat{U}^{\dagger}_{BS}$ and that beam splitters do not change the operator ordering, allowing us to directly act on the exponent of the normally ordered expression for the generating function.
    Furthermore, the beam-splitter transformation preserves the commutation relation among the mixed modes, which means that the tensor product of normal ordered operators that commute can be rewritten as the normal ordered tensor product of the functionals of $\hat{a}_k$ and $\hat{a}_{k'}$.
    Reordering the terms in the exponential with respect to the quadratic powers of creation and annihilation operators for the modes yields the final result as given in Eq. \eqref{formulafinale}.

\bibliography{biblio}
\end{document}